# Last Mile of Blockchains: RPC and Node-as-a-service

Zhongtang Luo
Purdue University
Email: luo401@purdue.edu

Rohan Murukutla
Supra
Email: r.murukutla@supraoracles.com

Aniket Kate
Purdue University / Supra
Email: aniket@purdue.edu

*Abstract*—While much research focuses on different methods to secure blockchain, information on the chain needs to be accessed by end-users to be useful. This position paper surveys different ways that end-users may access blockchains. We observe that between the two extremes of running a full node and fully utilizing a trusted third-party service, many solutions regarding light nodes are emerging. We analyze these solutions based on three basic properties of web communication: integrity, availability and privacy. We conclude that currently, the best way to access a blockchain while maintaining these three properties is still to run a full node. We consider it essential that future blockchain accessibility services should be built while considering these three expectations.

## I. Introduction

A blockchain is a group of nodes trying to maintain a publicly verifiable ledger of transactions protected from tampering. After its first proposal as Bitcoin [41] in 2009, blockchains have gained much popularity as financial instruments. The success of most well-known blockchains, such as Bitcoin and Ethereum [20] can largely be attributed to their decentralized nature and their capabilities to handle complex transactions with advanced cryptographic primitives. With currently a large market, blockchains have the potential to become a leading innovation in the finance and e-commerce sectors in the 21st century.

Much existing blockchain infrastructure research and innovation focus on secure and efficient ways for participating nodes to maintain the public ledger (known as a layer-1 protocol). However, we note that fetching information from a well-maintained ledger is also a non-trivial issue. Ensuring that blockchains stay accessible to every user is essential if blockchains are to play a bigger role in everyday life—a public ledger is, by definition, only useful as long as people can interact with it.

Given the recent growing public interest in blockchains, more users are interacting with blockchains every month, with as many as 875,000 transactions a day for popular blockchains such as Ethereum [20]. Trying to extract useful information from a flood of transactions requires powerful hardware for even a standard blockchain user who does not engage in profit-generating behavior (like mining), and it appears that the incentive mechanism in most blockchains is not equipped to deal with the issue.

In most blockchain protocols, such as Bitcoin and Ethereum, participating nodes (miners in proof-of-work systems and validators in proof-of-stake systems) are incentivized to maintain the ledger by earning a small fee when proposing a new block to the ledger. While participating nodes in these blockchain protocols are incentivized to collect transactions and broadcast raw blocks to ensure their profit, they are not incentivized to publish refined information about the ledger they maintain. Consider these two examples: (1) A user wants to know how many coins there are in their Bitcoin wallet. (2) A user wants to know the current status of a smart contract they want to execute.

Similar to these examples, most information users want to know is not directly related to the raw blocks that miners and validators publish. The users may entrust their assets fully to a third-party exchange and obtain information from the exchange, but the approach is significantly risky as evidenced by the recent FTX collpase [31]. Therefore, blockchain users need to devise a method to extract such information from blocks, either through some computation power that they own or by relying on a third-party service provider.

In this paper, we compare the two current methods in practice that users adopt to obtain refined information from blockchains. One method involves running a full node locally. This method is usually capable of extracting much information from the blockchain without external reliance. In some cases though, additional indices are kept to ensure a faster query, such as Ethereum archive nodes and Solana RPC nodes. The other method is to rely on a third-service provider. We notice the trend that as the cost of running a full node grows, especially in Ethereum, many users are turning to third-party services, which carry their own security risks. We observe that running a full node and fully utilizing a trusted third-party service represent two extremes regarding blockchain accessibility: one with minimal reliance and maximal hardware requirement and one with maximal reliance and minimal hardware requirement.

We also notice that when the required data on the blockchain is specific and limited, like when the users only care about one address that they own, specific wallet software is available to the users. While some of the wallet software queries an external node directly for data, others may download a portion of the blockchains and directly analyze it for the information the users care about. We categorize the software as specific light nodes. Meanwhile, we outline a recent trend in realizing a universal light node that can provide any information to users, not just one address [12], such as the one proposed by Ethereum [24]. Current research also includes so-called ultralight nodes, such as Flyclient [10], that step closer to relying on a third-party service that provides additional context for verification (also known as a prover in this context) than traditional light nodes that rely only on full nodes. Figure 1 shows a list of different methods to access blockchains.

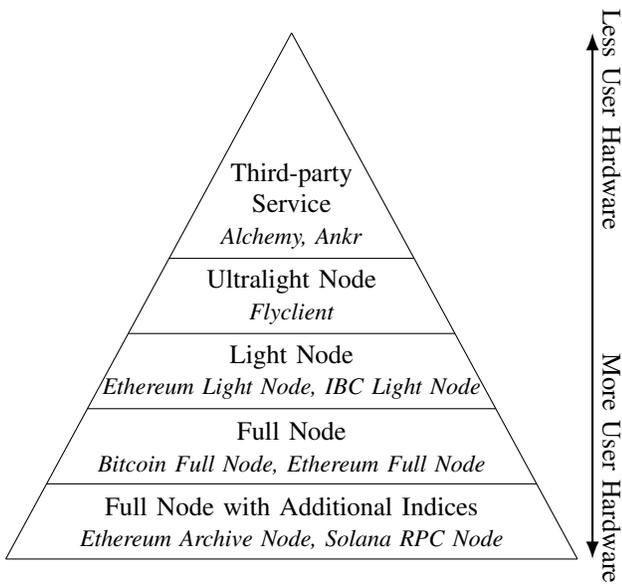

Fig. 1. Different methods to access blockchains and obtain useful information. Reliance on external resources decreases while reliance on local hardware power increases from top to bottom.

We think that amid many ways to access blockchains, it is still important to examine the basics of any web service: integrity, availability, and privacy. The blockchain itself is a technology of decentralization, yet too much reliance on a small number of third parties goes against this premise. Without a method to verify the correctness of data, integrity may be compromised if the said third parties provide incorrect data. With too much dependence on selected third parties to provide accessibility, availability may be impacted if they are unable or unwilling to provide service. Privacy is also a concern that needs addressing when querying any third party.

We examine existing protocols in the market and find that many of them may be flawed in at least one of these aspects. Currently, the best way to access a blockchain while maintaining integrity, availability and privacy is still to run a full node locally. We consider it important to pay attention to these three aspects when designing any protocol to interact with blockchains to ensure functionality.

## II. EXPECTATIONS OF BLOCKCHAIN ACCESSIBILITY

The process for a blockchain user to retrieve information from blockchains is similar to using any web service, such as browsing websites (retrieving information from a web server), checking e-mails (retrieving information from a mail server), etc. Therefore, the security expectations match the standard security triad: integrity, availability, and privacy.

### A. Integrity

For blockchain accessibility, integrity means that the user can obtain the correct information on blockchains under the given security model. Much of the existing research focuses on the integrity of the blockchains from the perspective of full nodes, guaranteeing integrity if the user has access to a trusted (e.g. self-owned) full node.

However, as more and more users turn to third-party service providers for information on blockchains, we notice that integrity may not be achieved if these service providers are untrusted. While some client protocols, such as Flyclient [10], ensure data integrity without much overhead, such techniques are not universally deployed on every blockchain or adopted by every service provider.

### B. Availability

For blockchain accessibility, availability means that the user can freely access information on blockchains. Most incidents on third-party service providers revolve around availability [36] [39], and at least one vulnerability that creates a denial-of-service has been studied [35].

These cases of empirical evidence suggest that fully relying on a single third-party service may create availability issues. While it is industry practice to diversify and use several third-party services as backups of each other, it may also help if the user can employ a solution that does not rely on a single service but utilizes a group of nodes to ensure maximum availability.

### C. Privacy

Privacy encompasses a broad range of cases available in this context: some users may want to conceal their queries from specific eavesdroppers to prevent frontrunning [17], while others may not want to be found using blockchains at all, with Bitcoin being illegal in some countries. These different concerns require drastically different security models that cannot be summarized in a nonspecific way.

Therefore, we devote our discussion of privacy in this paper strictly to the problems specific to blockchain accessibility rather than general Internet anonymity needs: an untrusted server of the network should have no information about the users' queries and the results.

While existing Internet protocols such as HTTPS largely realize end-to-end confidentiality between the user and the trusted server, we note that if resolving the user's query for information involves untrusted parties such as miners and third-party service providers, privacy is not trivially guaranteed and must be discussed on a case-by-case basis. For example, ConsenSys, the company that developed Infura and MetaMask, has claimed that when users use Infura as the default RPC provider in MetaMask, Infura will collect their IP address and Ethereum wallet address they send a transaction [15].

## III. APPROACH I: MAINTAIN A LEDGER LOCALLY - RUN A FULL NODE

For any user, the most straightforward way to access any blockchain is to maintain its ledger on their computer. A user can maintain a ledger by running a full node on their machine which constantly interacts with the whole blockchain network to keep track of the blocks on the ledger, even without participating in profit-generating activities like mining or staking. Such behavior is usually encouraged by blockchain protocols, as it is frequently cited to help propagate the peer-to-peer network and improve the security guarantee of the network [27].

Most blockchain services provide an optional interface within their node software that allows the user to fetch refined information from the node they own. The interface is usually implemented with the mechanism known as the Remote Procedure Call (RPC) [42]. In this request-response mechanism, the user's node serves as a server that accumulates information about the ledger through the network and answers the user's query (see Fig. 2) when they initiate one. Most RPC queries are transmitted as a JSON message between the client and the node. For example, in Bitcoin, a user can query the balance of their wallet by calling `getbalance`, sending the following JSON via HTTP to the node [8]:

```
{
    "jsonrpc": "1.0",
    "method": "getbalance",
    "params": ["*", 6]
}
```

The result from the node is a numerical value, indicating the amount of balance available in the loaded wallet.

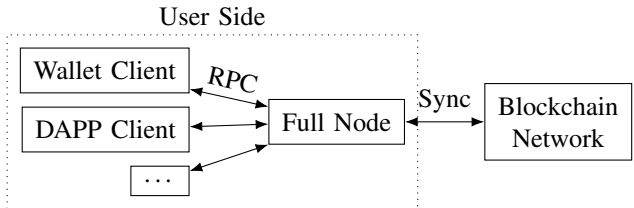

Fig. 2. Example of a user interacting with a full node they own via RPC. The full node constantly synchronizes with nodes in the blockchain network to keep track of information on the ledger. Different clients use RPC to interact with the node to query information about the blockchain.

### A. Cost Consideration

As the data on any blockchain continues to grow, running a node capable of accessing and storing all information on the ledger (known as a full node in most blockchains) is becoming increasingly more expensive. Official documentations and some unofficial tutorials outline the minimal system specification for Bitcoin [7], Ethereum [26], Solana [50], Zcash [21], Litecoin [5], Ripple [53], Dash [18], and Monero [14] (see Table I).

In particular, in blockchains based heavily on smart contracts, like Ethereum and Solana, full nodes have a high minimal system requirement: it is unlikely for every casual user to have a full-time server of 700GB of disk space, 8GB of RAM and 10Mbps of Internet connection dedicated to running an Ethereum full node.

Meanwhile, if the user wants to access more advanced information, they may need additional indexes built on top of a full node, which requires even more hardware power. For instance, an Ethereum archive node that can access all historical states takes more than 12TB disk space to run [26]. Solana's RPC nodes that provide RPC functionality also require a 16-core CPU in addition to more than 256GB RAM [50]. More accessibility functionalities such as subscribing to certain events may require the node to interface with the web socket, which further amplifies the hardware demand. Therefore, in most scenarios, casual users outsource at least part of the ledger to an external system and query the system when necessary.

### B. Open Questions

Traditional research on the cost of running nodes on blockchains usually focuses on the mining aspect [51] [54], since this is the only profit-generating activity of the proof-of-work blockchains. Because the cost of running a full node is negligible compared to the cost of mining equipment, it is usually overlooked.

While proof-of-stake systems do not have the problem of mining, the amount any validator has to stake to participate in the system usually outweighs the cost of running a full node, and the system-running cost is not considered heavily in the economics [28].

As a result, while the most secure way for any user to interact with the blockchain is to run their full node locally, it is not well-studied what the cost of keeping a full node online is. In particular, we note that some popular payment methods such as Zcash and Litecoin have little information on the hardware required to run a full node that synchronizes with the network. Studies are needed to determine the exact cost for users to access blockchains most securely.

On the flip side, there have been attempts to run full nodes of Bitcoin and Litecoin on Raspberry Pi 3, eliminating the setup of a traditional server entirely [37] [38]. It remains to be determined if such a method can be adopted to reduce the cost of running a full node.

## IV. APPROACH II: QUERY A THIRD-PARTY LEDGER - NODE-AS-A-SERVICE (NAAS)

If a user cannot afford to run a full node themselves, they must outsource the synchronization of blockchains to some third party. While typical users may be tempted to entrust their crypto coins and assets to an all-in-one cryptocurrency exchange, such as Crypto.com [16] and Binance [6], for exchange of easy accessibility, we consider the approach fundamentally flawed. Because the user has no access to the asset without the consent of the exchange, they cannot claim that they effectively own the asset: the recent collapse of FTX [31] reveals the risk of the exchange being insolvent and the user's asset being lost.

TABLE I
MINIMAL SYSTEM SPECIFICATION OF FULL NODES FOR VARIOUS BLOCKCHAINS. DASH MEANS UNSPECIFIED.

| Specs. | Bitcoin [7] | Ethereum [26] | Solana [50] | Zcash [21] | Litecoin [5] | Ripple [53] | Dash [18] | Monero [14] |
|---|---|---|---|---|---|---|---|---|
| CPU | - | 2-core | 12-core, 2.8GHz | - | - | 4-core | 2GHz | 2-core |
| RAM | 2GB | 8GB | 128GB | 8GB | 2GB | 16GB | 2GB | 4GB |
| Disk Space | 7GB | 700GB | 2TB | - | 25GB | 50GB | 40GB | 160GB |
| Network | 400Kbps | 10+Mbps | 300Mbps | - | - | 1Gbps | - | - |

TABLE II
POPULAR NODE-AS-A-SERVICE PROVIDERS, AS ADVERTISED BY ETHEREUM FOUNDATION [25]. PROTOCOLS SUPPORTED - NUMBER OF PROTOCOLS SUPPORTED BY THE SERVICE. FREE TIER - IF A FREE TIER IS AVAILABLE. MINIMUM PAYMENT - THE MINIMUM AMOUNT OF PAYMENT BEYOND THE FREE TIER (IF EXISTS), PER MONTH. CUSTOM API - IF A BLOCKCHAIN API DEVELOPED BY AND EXCLUSIVE TO THE SERVICE PROVIDER EXISTS.

| Features | Alchemy [2] | All That Node [3] | Ankr [4] | BlockDaemon [9] | Chainstack [11] | DataHub [19] |
|---|---|---|---|---|---|---|
| Protocols Supported | 7 | 24 | 18 | 60 | 20 | 4 |
| Free Tier | Yes | Yes | Yes | Yes | Yes | Yes |
| Minimum Payment | 49$ | ∗ | ⋆ | 400$ | 49$ | ∗ |
| Custom API | Yes | No | Yes | Yes | Yes | No |

| Features | GetBlock [29] | InfStones [32] | Infura [33] | Kaleido [34] | Moralis [40] | NOWNodes [43] |
|---|---|---|---|---|---|---|
| Protocols Supported | 40 | 50 | 11 | 5 | 2 | 50 |
| Free Tier | Yes | Yes | Yes | Yes | Yes | Yes |
| Minimum Payment | 29$ | 19$ | 50$ | ⋆ | 49$ | 20€ |
| Custom API | No | No | No | Yes | Yes | No |

| Features | Pocket Network [45] | QuickNode [46] | Rivet [47] | SenseiNode [48] | SettleMint [49] | Watchdata [52] |
|---|---|---|---|---|---|---|
| Protocols Supported | 15 | 14 | 1 | 8 | 6 | 5 |
| Free Tier | Yes | Yes | Yes | No | No | Yes |
| Minimum Payment | ∗ | 49$ | ⋆ | ∗ | 100€ | 49$ |
| Custom API | No | No | No | No | No | Yes |

| Features | ZMOK [55] |
|---|---|
| Protocols Supported | 1 |
| Free Tier | Yes |
| Minimum Payment | 54$ |
| Custom API | No |

∗ - Not publicly available. ⋆ - Pay as you go, i.e., the number of requests determines the charge.

Therefore, if the user wants to keep his private key and still maintain reasonable access to the blockchain, one approach is to purchase a third-party service that answers RPC queries, colloquially known as Node-as-a-service (NAAS) or an RPC node. The clients can interact with the third-party service just as they would with a full node (see Fig. 3).

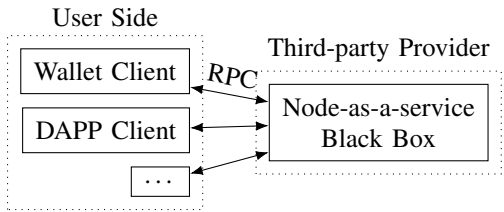

Fig. 3. Example of a user interacting with a third-party node-as-a-service provider. The clients communicate with the service in the same way as they communicate with a full node via RPC.

Many active NAAS providers exist in the field, including Alchemy, Ankr, BlockDaemon, etc. Table II provides a list of providers, as advertised by the Ethereum Foundation [25]. While we recognize that different providers offer services with different features and have different pricing plans, which makes an accurate comparison challenging, we pick four uniform measurements to understand the provider: the number of protocols supported, whether there is a free tier, the minimum amount the user needs to pay for a paid tier (per month), and whether or not the provider developed an exclusive API (such as one that handles the NFT) for the service in addition to traditional RPC calls.

*A. Pros and Cons*

The main benefit of using a NAAS service is that it is more affordable than running one's own full node because the service provider can build an infrastructure that handles a high workload efficiently. Many providers also have a free tier that allows a limited number of queries.

It is also noteworthy that most providers allow users to access more than one blockchain, which would traditionally require running one full node for each blockchain. Some providers offer exclusive pre-made APIs to handle common use cases, such as NFT. While using such API may reduce the program's portability, it increases the development efficiency.

The drawback of the approach is also significant: the security assumption of the model now depends on the service provider as a trusted party. Consider the case of Bitcoin's RPC call `getbalance`. The client has no way to verify the numerical result because RPC is not designed to ensure data integrity: every user is expected to run a node themselves. There is also no privacy guarantee when the service provider is untrusted: the service provider can see all parameters of an RPC call, thus increasing the risk associated, such as frontrunning [17].

Besides, any attack and incident of the service provider can impact users' access to blockchain information, thus affecting availability. Ankr was hit by a DNS attack in July 2022, which disrupted its users' access to two blockchains and placed their funds at risk [36]. Similarly, Infura was forced to censor Venezuelan users due to a US sanction in March 2022 [39]. These cases illustrate that a centralized service provider can become a single point of failure, and additional considerations are needed to ensure the accessibility of blockchains.

### B. Open Questions

Node-as-a-service providers form a competitive market: there are 19 providers advertised on the Ethereum Foundation website. However, there is little study on the market: who are the main users? Which service has a better cost-performance ratio for a specific type of user? How profitable can the providers be? An empirical study on this growing market is a challenging topic.

The design of the RPC protocol in this scenario can be particularly challenging since the current RPC protocol assumes the complete trust of the server. However, with the recent advancement of verification methods, such as Flyclient [10], a verifiable RPC protocol may be possible from a minimal-sized root of trust, solving the issue of integrity in the face of an untrusted service provider.

We also observe that it is standard industry practice to employ several services so that at least one is available at any time. Since many providers exist in the field, a unified structure that features a collaboration between providers may provide an improved availability guarantee and some degree of privacy and anonymity to users. For instance, Lava [13] is an ongoing proposal to design a decentralized RPC network blockchain that allows a group of NAAS providers to join together to provide services to users. The users will be charged in LAVA, a new type of coin specific to the network. Both NAAS providers and blockchain validators will stake LAVA to use the system. Unfortunately, Lava's whitepaper lacks a formal description of security, instead claiming that slashing the funds from bad actors can stop them from providing compromised information.

## V. APPROACH III: LIGHT NODE - EXTERNAL QUERY & LOCAL VERIFICATION

With the cost of running a full node continuously rising, the Ethereum community is seeking to develop a solution that can ensure data integrity and reduce system requirements. The result is the light node [23]: A light node is a server that only synchronizes limited information (such as block headers) with the blockchain network instead of block contents. When a light node encounters an RPC query, it pulls related contents from the blockchain network, verifies the contents with the synchronized information, and sends the result back to the client (see Fig. 4).

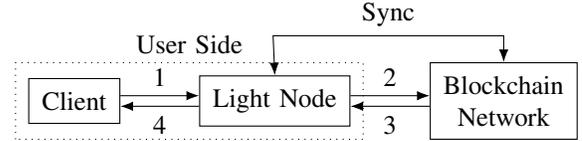

Fig. 4. Example of a user interacting with a light node with RPC. (1) The client sends an RPC query to the node. (2) The node asks the blockchain network for the necessary blocks to process the query. (3) The blockchain network delivers the blocks to the node. (4) The node verifies the blocks with the stored headers, processes the data, and sends the result back to the client.

Traditionally, a light node synchronizes all block headers with the network and pulls related blocks to answer the query when it receives one. However, the recent innovation of ultralight nodes, such as Flyclient [10], allows a light node to synchronize less data with the blockchain but still be able to verify data related to the query, provided that some node in the blockchain network is willing to compute a piece of proof for the correctness of the data.

When the required data on the blockchain is very specific and limited, like when the users only care about one address that they own, specific light nodes, in the form of wallet software, are also available to the users. For zero-knowledge-based blockchains (such as Zcash [21] and Monero [14]), the wallet software can synchronize with the blockchains by downloading and analyzing blocks one by one to extract all transactions related to the address, without actually storing the blocks. These specific light nodes can be much more efficient for new blockchain users because they only care about blocks that come after the creation of the address, so they do not have to analyze any data before the creation time.

There are also many studies on universal light nodes in the research area, but few have been deployed to real-world blockchains [12]. Some notable examples in active development include Flyclient [10] and IBC [30]. A light node is expected to ensure data integrity because it can check the block contents with the node's synchronized headers. However, a recent study shows that the security guarantee is weakened when the adversary controls the network [44].

### A. The Case of Ethereum

We notice that support for light nodes is minimal as of 2022 for Ethereum, one of the most popular blockchains [24]: there is only one execution client and no production-ready light clients on the consensus layer. The proper functioning of light nodes depends on their ability to pull blocks from full nodes in the network, but there is no incentive for full nodes to provide such service, and as a result, light nodes often fail to find peers.

Future development on light nodes focuses on allowing light nodes to relay information among themselves [22]: it is planned not to provide financial incentives for nodes, and the network is designed to work like BitTorrent and IPFS.

*B. Open Questions*

While significant engineering effort is still necessary before light clients become universal, there is still room for innovations to improve the client's performance. A recent SoK paper has suggested privacy-preserving, efficient resynchronization after offline, and cross-chain interoperability as potential future research directions [12].

Based on the premise that proving some statement to be true is more difficult than proving it to be false, some recent study also focuses on fraud proof [1]. The idea is that the light node does not need to verify the information it receives to be true, but can simply ask a few full nodes to make sure that it is not false.

We also notice that no matter the design of the light client, if it needs a full node to serve as a proofer for additional efforts beyond simply broadcasting a block, then incentives might be necessary to ensure that the full node renders the service. Toward that end, we think that the light node mechanism may merge with the node-as-a-service providers, with the providers serving the light nodes in a way that ensures integrity. In this sense, light nodes may evolve as a way to a verifiable RPC protocol.

## VI. Concluding Remarks

Ultimately, any blockchain accessibility solution has to decide on a split of responsibilities: either the user does most of the computation necessary or some third party does. A spectrum of solutions may be developed based on different types and needs of users, from a full node to a full third-party service. However, we consider that any solution must still adhere to reasonable security expectations: integrity, availability, and privacy.

Most blockchain research is based on the assumption that users run their local nodes. This means the fact that running a full node is still the best way to provide integrity, availability, and privacy. Indeed we surveyed third-party service providers in the field and noted the lack of a data integrity guarantee when the provider is not trusted. We also recalled events that render the providers unable to provide service. These facts mean that third-party service providers can and do fail, and thus relying on them can become risky.

At the same time, considering the high cost of running a full node locally, we believe that users should have more choice that does not impact the security guarantees. We observe the emergence of so-called ultralight nodes, which provides a way for third-party services to give a verifiable statement that requires little hardware power from the client.

Many open questions remain regarding the accessibility of blockchains, even regarding full-node solutions and third-party providers. Any number of light node protocols can also be built to serve different users' needs.

We reiterate that ensuring the accessibility of blockchains is paramount for the technology. If we believe that blockchains are a common good, then making them accessible to all in a secure and available manner is critical for the future development of Web3 technology. Any information is as useful as how it can be read, and such reasoning also applies to blockchains.